# Some simple rules for estimating reproduction numbers in the presence of reservoir exposure or imported cases

Angus McLure[1]*, Kathryn Glass[1]

[1] Research School of Population Health,
   Australian National University,
   62 Mills Rd, Acton,
   0200, ACT, Australia

* Corresponding author: angus.mclure@anu.edu.au

# Abstract

The basic reproduction number ($R_0$) is a threshold parameter for disease extinction or survival in isolated populations. However no human population is fully isolated from other human or animal populations. We use compartmental models to derive simple rules for the basic reproduction number for populations with local person-to-person transmission *and* exposure from some other source: either a reservoir exposure or imported cases. We introduce the idea of a *reservoir-driven* or *importation-driven* disease: diseases that would become extinct in the population of interest without reservoir exposure or imported cases (since $R_0 < 1$), but nevertheless may be sufficiently transmissible that many or most infections are acquired from humans in that population. We show that in the simplest case, $R_0 < 1$ if and only if the proportion of infections acquired from the external source exceeds the disease prevalence and explore how population heterogeneity and the interactions of multiple strains affect this rule. We apply these rules in two cases studies of *Clostridium difficile* infection and colonisation: *C. difficile* in the hospital setting accounting for imported cases, and *C. difficile* in the general human population accounting for exposure to animal reservoirs. We demonstrate that even the hospital-adapted, highly-transmissible NAP1/RT027 strain of *C. difficile* had a reproduction number <1 in a landmark study of hospitalised patients and therefore was sustained by colonised and infected admissions to the study hospital. We argue that *C. difficile* should be considered reservoir-driven if as little as 13.0% of transmission can be attributed to animal reservoirs.

# 1 Introduction

Many pathogens affecting humans circulate between humans and animals through contact, food or indirectly through common disease vectors in the environment. Other pathogens move across population boundaries due to the movement of people. In the absence of transmission from other populations or reservoirs, the basic reproduction number – the average number of secondary cases arising from each primary case in a susceptible population – determines whether a disease will die out or persist through ongoing person-to-person transmission. Effective interventions can interrupt transmission by reducing the basic reproduction number below 1 causing the disease to die out in that population. However, any reservoir exposure or imported cases will continue to replenish the infected population, and so a disease will die out in a population if and only if the basic reproduction number is <1 *and* all reservoir exposure and importation are avoided. There is a rich literature in meta-population models that capture the interactions of populations that introduce or reintroduce pathogens to one another (e.g. [1–4]). However, one often only has data or interest in a single population but needs to account for external sources of infections. It is in this context that we wish to generate some simple principles or rules for estimating the reproduction number.

Methods have been developed to estimate the human reproduction numbers of emerging zoonoses with limited person-to-person spread [5,6]. Others have developed methods to account for the often large proportion of imported cases at the beginning of new epidemics, which if excluded or treated as if locally acquired would overestimate the reproduction number [7]. Though the term 'elimination' has been defined in many different ways [8], reducing the local reproduction number below one is one measure of this progress [9], and is a necessary step towards global eradication. Methods have been developed to estimate the reproduction number that account for the potentially large proportion of imported cases in settings where progress is being made towards elimination [9]. However none of these methods account for susceptible depletion and so are restricted to diseases with very low prevalence [5,6,9] or calculate the *effective* reproduction number [7], which is not a threshold parameter for disease persistence. Starting with simple models and incorporating heterogeneity or multiple strains, we have derived simple rules for estimating the reproduction number in a population where the disease is at endemic equilibrium due to a combination of local person-to-person transmission *and* reservoir exposure or imported cases. Many of these rules only require knowledge of disease prevalence and the proportion of infections attributable to the external source. We have applied these rules in two case studies of *C. difficile* infections.

## 2 The SIS Model

We begin by adapting the simplest possible compartmental model: the standard SIS model with a homogeneous, well-mixed population without demographics. We include two sources of infection: (1) person-to-person transmission which is proportional to the number of people infected (rate: $\beta i$) and (2) constant transmission from some reservoir that does not depend on the number of people infected (rate: $f$). Person-to-person transmission could be through direct contact, or mediated via fomites, airborne droplets, water or food provided this transmission scales with the infectious population. For our purposes a reservoir is anywhere where the pathogen persists apart from the human population, for instance a population of wild animals or livestock animals that carry the disease. The disease in the human population can be described by a system of ODEs for the proportion of the population that is susceptible ($s$) and infected ($i$):

$$\begin{aligned} s' &= -\lambda(t)s + \gamma i \\ i' &= \lambda(t)s - \gamma i \end{aligned}$$

where $\lambda(t) = \beta i + f$ is the force of infection and $\gamma$ is the rate at which infected individuals recover.

Diseases that are acquired entirely from food or animals and diseases that are spread entirely by person-to-person transmission, are extreme cases of this model with $f = 0$ and $\beta = 0$ respectively. Many diseases lie between these two extremes. Almost all human cases of H7N9 avian influenza have been acquired from birds, but there has been some person-to-person transmission which is not enough to maintain endemic disease [10]. Meanwhile human-adapted seasonal influenza (H1N1, H3N2) are mainly transmitted to humans by other humans, though there are low frequency transmission events from animal reservoirs (e.g. [11]). Middle-eastern respiratory syndrome coronavirus sits somewhere in the middle of the spectrum with significant human-to-human and animal-to-human transmission [12].

The reproduction number for this simple model in the next-generation sense [13] is the same as for the standard SIS model ($\beta/\gamma$) but is a threshold parameter for extinction of the disease only when there is no transmission from the reservoir ($f = 0$), i.e. when the model reduces to the standard SIS model. Otherwise, the reservoir will continually replenish the infected population whatever the value of $R_0$. If there is no transmission from the reservoir we have the well-known relationship between the basic reproduction number ($R_0$) and the proportion of the population susceptible at equilibrium ($S$): $R_0 = 1/S$. The model parameters are difficult to measure directly and so we wish to estimate $R_0$ through observable quantities by generalising this rule. Let $I, S$ and $\Lambda = \beta I + f$ be the non-trivial (i.e. $I, \Lambda \neq 0$) equilibrium values of $i, s$ and $\lambda$. As equilibrium points of **1** they satisfy

$$\Lambda S = \gamma I$$

or equivalently

$$\gamma = S\frac{\Lambda}{I}.$$

Now the proportion of transmission that is from the reservoir at equilibrium, $\pi$, is

$$\pi = \frac{f}{\Lambda} = 1 - \frac{\beta I}{\Lambda},$$

which re-arranged for $\beta$ gives

$$\beta = (1-\pi)\frac{\Lambda}{I}.$$

Substituting these expressions for $\beta$ and $\gamma$ into the expression for the reproduction number we get

$$R_0 \equiv \frac{\beta}{\gamma} = \frac{1-\pi}{S}.$$

<div style="text-align: right;">**2**</div>

We can also write this in terms of the proportion infected (which is usually what is reported rather than the proportion susceptible).

$$R_0 = \frac{1-\pi}{1-I}.$$

<div style="text-align: right;">**3**</div>

These expressions simplify to $R_0 = 0$ if the disease is only acquired from the reservoir ($\beta = 0, \pi = 1$) or to $R_0 = \frac{1}{S} = \frac{1}{1-I}$ when all transmission is person-to-person ($f = 0, \pi = 0$). The general cases of these expressions lead to a simple rule for the reproduction number: $R_0 > 1$ if and only if $I > \pi$. The disease can be maintained by person-to-person transmission in the absence of reservoir exposure if and only if the prevalence exceeds the proportion of transmission from the reservoir.

This simple rule has surprising implications. For diseases with low prevalence (e.g. 2%), if a small but larger portion (e.g. 3%) of transmission is from the reservoir, then the disease cannot be sustained in the population by person-to-person transmission alone (since $R_0 = \frac{1-0.03}{1-0.02} < 1$). Preventing the small proportion of transmission from the reservoir (reducing $f$ and $\pi$ to 0) will cause the disease to become extinct in the population. Nevertheless, names like 'food-borne' or 'zoonotic' may be misleading for such diseases because the source of transmission is another human in most (e.g. 97%) individual infections. Instead we call these diseases *reservoir-driven*. We define the *reservoir-driven threshold* as the minimum proportion of transmission which must be from the reservoir for the disease to be considered reservoir-driven ($I$ in our simple SIS model).

The rest of this article will consider variants and extensions of the simple SIS model to demonstrate which assumptions do and do not affect the above expressions for the reproduction number and reservoir-driven threshold. We will also show that an equivalent rule of thumb and threshold exists when a disease is driven by imported cases due to travel or immigration. We will then consider how this rule of thumb can be applied to case studies in real diseases.

## 3 Simple Extensions of the SIS Model

### 3.1 Births and Deaths

Simple demographics does not change our rule for the reproduction number. A modified model including deaths from both classes at rate $\delta$ and births that balance deaths is described by the equations

$$\begin{aligned} s' &= -\lambda(t)s + \gamma i - \delta s + \delta \\ i' &= \lambda(t)s - \gamma i - \delta i \end{aligned},$$

where $\lambda(t) = \beta i + f$ is the force of infection. In this model $R_0 = \frac{\beta}{\gamma+\delta}$. Let $I, S$ and $\Lambda = \beta I + f$ be the non-trivial (i.e. $I, \Lambda \neq 0$) equilibrium values of $i, s$ and $\lambda$. Then

$$\Lambda S = (\gamma + \delta)I,$$

or equivalently

$$\gamma + \delta = S\frac{\Lambda}{I}.$$

The force of infection terms are the same as for our original model so again we have $\beta = (1 - \pi)\frac{\Lambda}{I}$. Substituting this into the expression for the reproduction number we get the same result as **2** and **3**:

$$R_0 \equiv \frac{\beta}{\gamma + \delta} = \frac{1 - \pi}{S} = \frac{1 - \pi}{1 - I}$$

and the reservoir-driven threshold is still $I$. We have assumed that the death rates are the same for infected and susceptible persons, but it is simple to show that a higher (or lower) death rate for infected individuals does not affect the reasoning.

## 3.2 Recovered Classes and Other Common Extensions

The simplest SIR model without birth and deaths or waning immunity does not have an endemic equilibrium point so our method for estimating the reproduction number is not applicable to these models. Instead, consider the SIR model with births and deaths:

$$\begin{aligned} s' &= -\lambda(t)s - \delta s + \delta \\ i' &= \lambda(t)s - \gamma i - \delta i \\ r' &= \gamma i - \delta r \end{aligned},$$

where $\lambda(t) = \beta i + f$ the force of infection. Note that adding the recovered class to the SIS model with births and deaths does not change the reproduction number, the equation governing the number of infected individuals or the force of infection and so the reasoning is identical to previous section. However, since there are more than two classes, $S + I \neq 1$. Therefore **3** does not hold but instead,

$$R_0 = \frac{1 - \pi}{S} = \frac{1 - \pi}{1 - (I + R)}.$$

The reservoir-driven threshold here is $I + R$, i.e. the disease can be sustained by person-to-person transmission in the absence of reservoir exposure if and only if the proportion of transmission which is due to reservoir exposure is less than the total proportion of people infected or immune/recovered.

The same reasoning can be used for models with waning immunity, vaccination, or latent/exposed classes. Since these modifications do not affect the equations governing the number of infected individuals or the force of infection, equation **2** still holds and therefore the reservoir-driven threshold is $1 - S$ in all these cases. For diseases with comprehensive vaccination programs (or common diseases with lifelong immunity), almost all the population can be immune (e.g. 95%) and the proportion susceptible very low. If reservoir exposure accounts for nearly all cases but is still less than the reservoir-driven threshold (e.g. 90%), the disease could be sustained by person-to-person transmission alone if reservoir exposure was eliminated (since $R_0 = \frac{1-0.90}{1-0.95} > 1$) and so eliminating exposure to the reservoir would not eliminate the disease from the human population.

## 3.3 Imported Cases

Analogous rules can be derived for settings where some infections are acquired locally and others are imported through immigration or those returning from travel. We assume that susceptible and infected individuals emigrate/leave at the same rate $\delta$, that immigration balances emigration and that a proportion $p$ of those entering the population are infected. The governing equations are

$$\begin{aligned} s' &= -\lambda(t)s + \gamma i - \delta s + (1-p)\delta \\ i' &= \lambda(t)s - \gamma i - \delta i + p\delta \end{aligned}$$

where $\lambda(t) = \beta i$ is the force of infection. Again, $R_0 = \beta/(\gamma + \delta)$ but $R_0$ is not by itself a threshold parameter for disease extinction because the continuous immigration of new infected individuals will sustain the disease (unless $p\delta = 0$). The equilibrium proportion infected ($I$), proportion susceptible ($S$) and force of colonisation ($\Lambda = \beta I$) satisfy

$$\Lambda S + p\delta = (\gamma + \delta)I,$$

or equivalently

$$\gamma + \delta = \frac{\Lambda S + p\delta}{I}.$$

Meanwhile the proportion of new cases that are imported, $q$, is

$$q = \frac{p\delta}{\Lambda S + p\delta} = 1 - \frac{\Lambda S}{\Lambda S + p\delta} = 1 - \frac{\beta I S}{\Lambda S + p\delta}$$

which we can rearrange for the transmission parameter giving

$$\beta = \frac{(1-q)(\Lambda S + p\delta)}{IS}.$$

Therefore, we can write the reproduction number as

$$R_0 \equiv \frac{\beta}{\gamma + \delta} = \frac{1-q}{S} = \frac{1-q}{1-I}.$$

These expressions lead to simple rules for the reproduction number analogous to those derived for diseased reservoir exposure. $R_0 > 1$ if and only if $I > q$. That is, in this simple model, the disease can be sustained without importation by local transmission if and only if the prevalence exceeds the proportion of new cases that are imported through migration or travel. By analogy to the reservoir exposure model, we call this threshold the *importation-driven* threshold. This analogy still holds when heterogeneity or multiple strains are incorporated into these models – extensions we consider in sections 4 and 5.

## 4 Heterogeneity

It is known that accounting for population heterogeneity tends to increase estimates of reproduction numbers [14]. Therefore, we might expect that introducing heterogeneity into models with reservoir exposure will increase the reservoir-driven threshold. Consider a general SIS model with separable mixing in a heterogeneous population consisting of people of different $x$-types with susceptibility $\alpha(x)$, transmission parameter $\beta(x)$ and mean infectious period $1/\gamma(x)$, distributed according to the probability density function $g(x)$. Then

$$\begin{aligned} s_t(x,t) &= -\lambda(t)\alpha(x)s(x,t) + \gamma(x)i(x,t) \\ i_t(x,t) &= \lambda(t)\alpha(x)s(x,t) - \gamma(x)i(x,t) \end{aligned}$$

and

$$s(x,t) + i(x,t) = g(x),$$

where $\lambda(t) = f + \int \beta(x)i(x,t)dx$. For this model, $R_0 = \int \beta(x)/\gamma(x)\alpha(x)g(x)dx$ [15], but as before $R_0$ is only a threshold parameter for disease extinction if $f = 0$. Let $I$ and $S$ be the non-trivial equilibrium distributions of $i, s$ and $\Lambda = \int \beta(x)I(x)dx + f$ the equilibrium value of $\lambda$. As equilibrium points they satisfy

$$\Lambda S(x)\alpha(x) = \gamma(x)I(x),$$

or equivalently,

$$\frac{S(x)\alpha(x)}{\gamma(x)} = \frac{I(x)}{\Lambda}.$$

<div style="text-align:right">4</div>

At equilibrium, the proportion of infections acquired from the reservoir, which is the proportion of force of infection attributable to the reservoir, is

$$\pi = \frac{f}{\Lambda} = 1 - \int \frac{\beta(x)I(x)}{\Lambda}dx.$$

Substituting 4 into the above gives

$$\pi = 1 - \int \frac{\beta(x)\alpha(x)S(x)}{\gamma(x)}dx.$$

If we let $\rho = \beta\alpha/\gamma$ we can write the reproduction number and the proportion of infections from the reservoir in simpler terms

$$\begin{aligned} R_0 &= \int \rho(x)g(x)dx \\ &= \overline{\rho} \end{aligned}$$

where $\overline{\rho}$ is the mean value of $\rho$ across the population and

$$\begin{aligned} \pi &= 1 - \int \rho(x)S(x)dx \\ &= 1 - \boldsymbol{S} \int \rho(x)\frac{S(x)}{\boldsymbol{S}}dx \\ &= 1 - \boldsymbol{S}\,\overline{\rho_S} \end{aligned}$$

where $\boldsymbol{S} := \int S(x)dx$ is the total susceptible population and $\overline{\rho_S}$ is the mean value of $\rho$ across the susceptible population. Therefore

$$R_0 = \frac{1-\pi}{\boldsymbol{S}\,\frac{\overline{\rho_S}}{\overline{\rho}}}.$$

By similar reasoning one can show that

$$R_0 = \frac{1-\pi}{1 - I\frac{\int \rho(x)\frac{I(x)}{I}dx}{\int \rho(x)g(x)dx}} = \frac{1-\pi}{1 - I\frac{\overline{\rho_I}}{\overline{\rho}}}$$

where $I := \int I(x)dx$ is the proportion of the whole population that is infected and $\overline{\rho_I}$ is the mean value of $\rho$ across the infected population. The quantity we want to estimate, $R_0$, appears as $\overline{\rho}$ in the right-hand sides of each equation and the quantities $\overline{\rho_I}$ and $\overline{\rho_S}$ are unlikely to be known, so this does not provide a practical way to estimate the reproduction number. However, these statements provide some insight into how heterogeneity can affect our estimates of the reproduction number or reservoir-driven threshold. The rule of thumb is similar to the rule for a homogenous population: $R_0 > 1$ if and only if $I\,\overline{\rho_I}/\overline{\rho} > \pi$, i.e. the reservoir-driven threshold is $I\,\overline{\rho_I}/\overline{\rho}$. If those who are infected have higher-than-average (or lower-than-average) $\rho$, then accounting for this heterogeneity increases (or decreases) the reservoir-driven threshold. We derive simple expressions for the value of $\overline{\rho_I}/\overline{\rho}$ under some specific assumptions.

## 4.1 Variable Susceptibility or Infectious Period

If we assume that the infectiousness ($\beta$) is fixed but the product of susceptibility and length of infectious period ($\phi := \alpha/\gamma$) is heterogeneous, then the reservoir-driven threshold is always higher than for a homogenous population. Consider the ratio $\overline{\rho_I}/\overline{\rho}$

$$\frac{\overline{\rho_I}}{\overline{\rho}} = \frac{\int \frac{\beta\alpha(x)}{\gamma(x)}\frac{I(x)}{I}dx}{\int \frac{\beta\alpha(x)}{\gamma(x)}g(x)dx} = \frac{\int \phi(x)\frac{I(x)}{I}dx}{\int \phi(x)g(x)dx} = \frac{\overline{\phi_I}}{\overline{\phi}},$$

where $\overline{\phi}$ and $\overline{\phi_I}$ are the mean values of $\phi$ across the whole population and across the infected portion of the population respectively. Now we can rearrange **4** in terms of the odds of infection of an individual of type $x$

$$\frac{I(x)}{S(x)} = \frac{\Lambda\alpha(x)}{\gamma(x)} = \Lambda\phi(x).$$

Since the odds of infection for an individual of type $x$ is proportional to $\phi(x)$, individuals with high $\phi$ (i.e. more susceptible individuals or individuals with longer infectious periods) will be over-represented in the infected portion of the population at equilibrium. Therefore $\overline{\phi_I} \geq \overline{\phi}$ and so the reservoir-driven threshold is at least as high as for a homogenous population.

If the prevalence is low for people of all $x$-types (i.e. $S(x) \approx g(x)$) there is a simple approximation for the reservoir-driven threshold. We can rearrange **4** to get

$$I(x) = S(x)\Lambda\phi(x) \approx g(x)\Lambda\phi(x).$$

and so

$$I = \int I(x)dx \approx \Lambda \int \phi(x)g(x)dx = \Lambda\overline{\phi}$$

and

$$\overline{\phi_I} = \int \phi(x)\frac{I(x)}{I}dx \approx \frac{1}{\overline{\phi}}\int \phi(x)^2 g(x)dx.$$

If the population variance of $\phi$ is $\sigma^2 := \int (\phi(x) - \overline{\phi})^2 g(x)dx$ the ratio can be written approximately as

$$\frac{\overline{\phi_I}}{\overline{\phi}} \approx \frac{1}{\overline{\phi}^2}\int \phi(x)^2 g(x)dx = 1 + \frac{\sigma^2}{\overline{\phi}^2}$$

and the reproduction number is

$$R_0 \approx \frac{1-\pi}{1 - I\left(1 + \frac{\sigma^2}{\overline{\phi}^2}\right)}.$$

When there is no heterogeneity in $\phi$ (i.e. when $\sigma^2 = 0$) this simplifies to the result for the homogenous SIS model. The larger the variance for a given mean, the greater the basic reproduction number and the higher the reservoir driven-threshold, $I(1 + \sigma^2/\overline{\phi}^2)$. For example if $\phi(x)$ and $g(x)$ are such that the distribution of $\phi$ across the population is gamma with mean $\mu$ and shape parameter $k$ (a convenient and often used assumption [14]), then the reservoir-driven threshold is approximately $I\left(1 + \frac{1}{k}\right)$ (**Figure 1**).

If the $x$ type of an individual corresponds to some easily determined risk class – for instance if $x$ denotes gender or smoker status – then the proportion of people in each class, $g(x)$, and the odds of infection within each group, $I(x)/S(x)$, may be known. Since the odds of infection is proportional to $\phi$, we can express $\overline{\phi_I}/\overline{\phi}$ and $R_0$ in terms of these observed quantities:

$$\frac{\overline{\phi_I}}{\overline{\phi}} = \frac{\int \phi(x)\frac{I(x)}{I}dx}{\int \phi(x)g(x)dx} = \frac{\int \frac{I(x)}{S(x)}\frac{I(x)}{I}dx}{\int \frac{I(x)}{S(x)}g(x)dx}$$

and

$$R_0 = \frac{1-\pi}{1 - I\frac{\overline{\phi_I}}{\overline{\phi}}} = \frac{1-\pi}{1 - I\frac{\int \frac{I(x)}{S(x)}\frac{I(x)}{I}dx}{\int \frac{I(x)}{S(x)}g(x)dx}}.$$

## 4.2 Variable Infectiousness

If infectiousness ($\beta$) is heterogeneous, but the product of susceptibility and length of the infectious period ($\phi := a/\gamma$) is fixed, then the reservoir-driven threshold is the same as for a homogenous population. Consider the ratio $\overline{\rho_I}/\overline{\rho}$ which can be simplified as

$$\frac{\overline{\rho_I}}{\overline{\rho}} = \frac{\int \beta(x)\phi\frac{I(x)}{I}dx}{\int \beta(x)\phi g(x)dx} = \frac{\int \beta(x)\frac{I(x)}{I}dx}{\int \beta(x)g(x)dx} = \frac{\overline{\beta_I}}{\overline{\beta}},$$

where $\bar{\beta}$ and $\bar{\beta_I}$ are the mean values of $\beta$ across the whole population and across the infected portion of the population respectively. Now by **6**, if $\phi$ is constant across the population the odds of infection at equilibrium is the same for people of every $x$-type, i.e. independent of their infectiousness. Therefore, the mean infectiousness amongst the infected population is the same as the mean infectiousness across the whole population and $\bar{\beta_I}/\bar{\beta} = \bar{\rho_I}/\bar{\rho} = 1$. Equations **5** then simplifies to

$$R_0 = \frac{1-\pi}{1-I}$$

which is the same as the result for a homogenous population.

Heterogeneous infectiousness will affect the reservoir-driven threshold in a population which is also heterogeneous with respect to susceptibility or infectious period. If those who are more likely to be in the infected class (high $\phi$) are less infectious (low $\beta$), this will reduce the reservoir-driven threshold relative to homogeneous infectiousness but heterogeneous susceptibility and infectious period. As a simple example of this, assume that $\beta(x) = 1/\phi(x)$. Then $\rho(x) = 1$, $\bar{\rho} = \bar{\rho_I} = 1$ and the reservoir-driven threshold is simply $I$, less than what it would be if $\beta$ were constant across the population. On the other hand, if those who are more likely to be colonised (high $\phi$) are also more infectious (high $\beta$), the reservoir-driven threshold will increase relative to homogeneous infectiousness but heterogeneous susceptibility and infectious period. As another simple example consider the proportional mixing assumption, i.e. $\beta \propto \phi$. In this case $\rho \propto \phi^2$ and so

$$\frac{\bar{\rho_I}}{\bar{\rho}} = \frac{\int \phi(x)^2 \frac{I(x)}{I} dx}{\int \phi(x)^2 g(x) dx} \ .$$

When the prevalence is low for people of all $x$-types (i.e. $S(x) \approx g(x)$) we can use the same reasoning as in the previous section to approximate this ratio as

$$\frac{\bar{\rho_I}}{\bar{\rho}} \approx \frac{\int \phi(x)^3 g(x) dx}{\bar{\phi} \int \phi(x)^2 g(x) dx} = \frac{\nu}{\bar{\phi}(\bar{\phi}^2 + \sigma^2)}$$

and the reproduction number by

$$R_0 \approx \frac{1-\pi}{1 - I\left(\frac{\nu}{\bar{\phi}(\bar{\phi}^2 + \sigma^2)}\right)} \ .$$

where $\nu$ is the third raw moment of $\phi$ across the population. If for example, $\phi(x)$ and $g(x)$ are such that $\phi$ is gamma distributed with shape parameter $k$ then the reservoir-driven threshold is approximately $I(1 + \frac{2}{k})$, which is higher than if $\beta$ were homogeneous. **Figure 1** summarises how the reservoir-driven threshold changes for different types of heterogeneity explored so far.

## 4.3 Variable Exposure to Reservoir

Heterogeneous exposure to the reservoir in an otherwise homogeneous population does not change the reservoir-driven threshold. Consider an SIS model where the population consists of people of type $x$ distributed according to $g(x)$ each with their own level of exposure to reservoir $f(x)$. Then the differential equations governing the system are

$$\begin{aligned} s_t(x,t) &= -\lambda(x,t)s(x,t) + \gamma i(x,t) \\ i_t(x,t) &= \lambda(x,t)s(x,t) - \gamma i(x,t) \end{aligned}$$

and

$$s(x,t) + i(x,t) = g(x),$$

where $\lambda(x,t) = \beta \int i(\xi,t)d\xi + f(x)$ is the force of infection acting on individuals of type $x$. The basic reproduction number for this model is $\beta/\gamma$. Let $I$ and $S$ be the non-trivial equilibrium distributions of $i, s$ (i.e. $I \neq 0$), $\boldsymbol{I}$ and $\boldsymbol{S}$ be the total number of people infected and susceptible at equilibrium and $\Lambda(x) = \beta \int I(x)dx + f(x) = \beta \boldsymbol{I} + f(x)$ be the equilibrium force of infection. As equilibrium points they satisfy

$$\Lambda(x)S(x) = \gamma I(x).$$

The proportion of transmission that is acquired from the reservoir is then

$$\begin{aligned} \pi &= \frac{\int f(x)S(x)dx}{\int \Lambda(x)S(x)dx} \\ &= 1 - \frac{\int \beta \boldsymbol{I} S(x)dx}{\int \Lambda(x)S(x)dx} \\ &= 1 - \frac{\int \beta \boldsymbol{I} S(x)dx}{\int \gamma I(x)dx} \\ &= 1 - \frac{\beta}{\gamma}\boldsymbol{S} = 1 - R_0 \boldsymbol{S} \end{aligned}$$

and consequently

$$R_0 = \frac{1-\pi}{\boldsymbol{S}} = \frac{1-\pi}{1-\boldsymbol{I}}$$

leaving the reservoir-driven threshold unchanged. However, interactions with additional heterogeneities will affect the reservoir-driven threshold. Consider the case where both reservoir exposure ($f$) and the person-to-person transmission rate ($\beta$) depend on the $x$-state. In this case the equilibrium force of infection is $\Lambda(x) = \int \beta(\xi)I(\xi)d\xi + f(x)$, the reproduction number is $R_0 = \int g(\xi)\beta(\xi)/\gamma\, d\xi = \overline{\beta}/\gamma$ where $\overline{\beta}$ is the mean value of $\beta$ in the population. The proportion of infections that are acquired from the reservoir is

$$\begin{aligned} \pi &= \frac{\int S(x)f(x)}{\int S(x)\Lambda(x)dx} \\ &= 1 - \frac{\int S(x)\int I(\xi)\beta(\xi)d\xi dx}{\int S(x)\Lambda(x)dx} \\ &= 1 - \int I(\xi)\beta(\xi)d\xi \frac{\int S(x)dx}{\int I(x)\gamma dx} \\ &= 1 - \int \frac{I(\xi)}{\boldsymbol{I}}\beta(\xi)d\xi \frac{\boldsymbol{S}}{\gamma} \\ &= 1 - \frac{\overline{\beta_I}}{\gamma}\boldsymbol{S} = 1 - R_0 \frac{\overline{\beta_I}}{\overline{\beta}}\boldsymbol{S}, \end{aligned}$$

where $\overline{\beta_I}$ is the mean value of $\beta$ in the infected population. Therefore

$$R_0 = \frac{1-\pi}{S\frac{\overline{\beta_I}}{\overline{\beta}}} = \frac{1-\pi}{(1-I)\frac{\overline{\beta_I}}{\overline{\beta}}}.$$

Those that have greater exposure to the reservoir are more likely to be infected and so will have a disproportionally large effect on $\overline{\beta_I}$. If those with more exposure to the reservoir are also on more infectious then $\overline{\beta_I} > \overline{\beta}$ and the reservoir-driven exposure is lower, and conversely if those with more exposure to the reservoir also less infectious then $\overline{\beta_I} < \overline{\beta}$ and the reservoir-driven threshold is higher (**Figure 2**). Note that this is opposite to the relationship for heterogeneous $\beta$ and heterogeneous $\phi$ (**Figure 1**).

## 5 Multiple Strains

There is frequently more than one strain of a pathogen co-circulating within human populations and the dynamics of multi-strain interactions have been modelled extensively (e.g. [16–20]). In the few simple multi-strain models we consider, accounting for host competition increases the reservoir driven threshold for each strain compared to the single strain model. Consider a simple competitive multi-strain extension of our basic SIS model with reservoir exposure. Each strain has its own transmission parameter ($\beta_k$), recovery rate ($\gamma_k$) and reservoir exposure rate ($f_k$). We assume that infection with one strain prevents infection from all other strains for the duration of the infection. With $n$ strains the $n+1$ equations governing this system are

$$s' = -\sum_{k=1}^{n} \lambda_k(t)s + \sum_{k=1}^{n} \gamma_k i_k$$
$$i_k' = \lambda_k(t)s - \gamma_k i_k, \quad k = 1, \ldots, n$$

where $\lambda_k(t) = \beta_k i_k(t) + f_k$ is the force of infection for each strain. Each strain has its own basic reproduction number in a fully susceptible population: $R_0^k = \beta_k/\gamma_k$. Here, $R_0^k$ are not threshold parameters for strain extinction because reservoir exposure will cause the disease to persist *and* strain competition for hosts may cause a strain without reservoir exposure to die out even if that strain's reproduction number exceeds one. Let $S$, be the equilibrium number of susceptible people at the non-trivial equilibrium where the number of people infected with each strain ($I_k$) and the force of colonisation for each strain ($\Lambda_k$) are non-zero. For each strain we have the following relation

$$\Lambda_k S = \gamma_k I_k,$$

or equivalently

$$\gamma_k = \frac{\Lambda_k S}{I_k}.$$

The proportion of transmission of strain $k$ that is from the reservoir is

$$\pi_k = \frac{f_k}{\Lambda_k} = 1 - \frac{\beta_k I_k}{\Lambda_k}.$$

Rearranging for $\beta_k$:

$$\beta_k = (1 - \pi_k)\frac{\Lambda_k}{I_k}.$$

We can re-write the basic reproduction number for strain $k$ as

$$R_0^k \equiv \frac{\beta_k}{\gamma_k} = \frac{1 - \pi_k}{S} = \frac{1 - \pi_k}{1 - \sum_{j=1}^n I_j}.$$

Consequently $R_0^k < 1$ if $\pi_k > \sum_{j=1}^n I_j$. It follows that a given strain cannot persist without reservoir exposure if the proportion of transmission of that strain due to reservoir-exposure is more than the total prevalence of *all* strains.

We also want to account for strain competition which can lead to the extinction of strains that would otherwise persist in a population. Therefore, we consider the invasion reproduction number for each strain, i.e. not the reproduction number in a fully susceptible population, but in a population at endemic equilibrium with all the *other* strains. Consider the equilibrium point without any infections of strain $k$ that exists if there is no reservoir exposure for strain $k$ (i.e. $f_k = 0$). Let $S^k, I_1^k, \ldots, I_n^k$, be the equilibrium values of $s, i_1, \ldots, i_n$ when $f_k = 0$, such that $I_k^k = 0$ and $I_j^k > 0$ if $j \neq k$. The invasion reproduction number for strain $k$ is then $R_{Invasion}^k = R_0^k S^k$. It is possible to calculate $S^k$ in terms of $\pi_1, \ldots, \pi_n$ and $I_1, \ldots, I_n$ but the exact form is cumbersome (even for the $n = 2$ case) so instead we consider a simple bound. Consider that the equilibrium proportion of each strain other than $k$ will certainly not decrease in the absence of the competition with strain $k$, i.e. $I_j^k \geq I_j$ for $j \neq k$. Consequently $S^k \leq S + I_k$ since

$$S^k = 1 - \sum_{\substack{j=1 \\ j \neq k}}^n I_j^k \leq 1 - \sum_{\substack{j=1 \\ j \neq k}}^n I_j = 1 + I_k - \sum_{j=1}^n I_j = S + I_k.$$

We can bound the invasion reproduction number for strain $k$ by

$$\begin{aligned} R_{Invasion}^k &= R_0^k S^k \\ &\leq \frac{1 - \pi_k}{S}(S + I_k) \\ &= \frac{1 - \pi_k}{1 - \frac{I_k}{S + I_k}}. \end{aligned}$$

Consequently $\frac{I_k}{S+I_k}$ is an upper bound for the reservoir-driven threshold in the presence of other strains since $R_{Invasion}^k < 1$ whenever $\pi_k > \frac{I_k}{S+I_k}$.

Our simple competitive model assumes complete exclusion, but in reality, strains are unlikely to completely exclude one another. If one allows for the possibility of coinfections or superinfection, assuming that persons infected with strains other than strain $k$ ($I_{\neg k}$) are $a_k$ times as susceptible to infection with strain $k$ as those not infected with any strain ($S$) and that coinfecting/superinfecting strains do not affect infectiousness or infectious period for the infecting strains, then at endemic equilibrium

$$\Lambda_k(S + a_k I_{\neg k}) = \gamma_k I_k$$

where $I_k$ is the proportion of people infected with strain $k$ (who may also be infected with other strains) and $\Lambda_k = \beta_k I_k + f_k$. One can use the same reasoning as above to show that

$$R_0^k \equiv \frac{\beta_k}{\gamma_k} = \frac{1-\pi_k}{S + a_k I_{\neg k}} = \frac{1-\pi_k}{1 - I_k + (1-a_k)I_{\neg k}}$$

and

$$R_{Invasion}^k \leq \frac{1-\pi_k}{1 - \frac{I_k}{S + a_k I_{\neg k} + I_k}} = \frac{1-\pi_k}{1 - \frac{I_k}{1-(1-a_k)I_{\neg k}}}.$$

and so $\frac{I_k}{S+a_k I_{\neg k}+I_k} = \frac{I_k}{1-(1-a_k)I_{\neg k}}$ is an upper bound for the reservoir-driven threshold. If $a_k = 0$, this reduces to the case of complete exclusion we considered above. If $a_k = 1$, that is if infection with another strain neither prevents nor predisposes a patient to infection with strain $k$, then the reservoir driven threshold and reproduction number are the same for as for a model with only a single strain. In general, the greater the exclusion against strain $k$ (i.e. as $a_k \to 0$), the higher the reservoir-driven threshold and reproduction number. Consequently the case of complete exclusion is an upper bound for these quantities in these simple models.

## 6 Case Study: *Clostridium difficile*

*Clostridium difficile* is a bacterium that colonises the intestines of many mammals including humans and livestock [21]. Most human hosts do not have symptoms despite being colonised. Colonisation is typically transient, lasting approximately one month in adults [22], due to competition and interactions with other intestinal flora [23]. Disruption of the gut flora, often caused by consumption of antibiotics or proton-pump-inhibitors, allows *C. difficile* to proliferate in large numbers [23]. Toxigenic strains of *C. difficile* then produce a number of toxins that can cause diarrhoea which is often severe and sometimes life-threatening [24]. A robust immune response to these toxins is able to neutralise their effect [25] and most of the population have anti-toxin antibodies starting at a young age [26]. Immune responses protect against symptoms but not protect against colonisation [27]. Asymptomatically colonised carriers are also infectious [28] while animal models have shown that disruption of gut flora, even in the absence of symptoms, increases spore shedding and infectiousness [29].

Since immunity does not prevent colonisation or infectiousness, a simple SIS model is an appropriate starting point for *C. difficile*, provided we identify the I-class with all *C. difficile* positive individuals (not just those with symptoms). We will use variations on the SIS model below to determine whether *C. difficile* is importation-driven in a hospital setting, and calculate the reservoir-driven threshold for *C. difficile* for the human population as a whole (where animals are the reservoir).

### 6.1 *C. difficile* in Hospitals

Historically, *C. difficile* has been of most concern and thus most studied in hospitalised patients where it complicates the care of many initially hospitalised for other conditions [30]. However, there is growing recognition of community-acquired cases that manifest during hospital stay. Since *C. difficile* is consistently present in many hospitals, it has been assumed that *C. difficile* is endemic in these settings and is responsible for many cases in the community.

If we begin by modelling *C. difficile* in hospitals as a (homogeneous) SIS model with very high rates of migration (hospital admission and discharge) then we can estimate the reproduction number using the method outlined in Section 3.3. In words, we will estimate the within-hospital reproduction number as

$$R_0^{\text{Hospital}} = \frac{1 - \text{Proportion of colonisations and infections acquired prior to admission}}{1 - \text{Prevalence of colonisation and infection}}.$$

One study of colonisation and infections in hospitalised patients found 184 patients colonised at admission, and another 240 patients that acquired colonisation or infection after admission [27]. They identified an additional 60 or so patients that developed a CDI within 72 hours of admission who were therefore deemed to have been exposed prior to admission. Thus, the proportion of *C. difficile* positive patients that acquired the pathogen prior to admission was approximately 50%.

In the same study 528/5422 patients were colonised or developed an infection for part of their hospital stay. Some patients were excluded from their analysis (mostly for missing data) leaving 424/4143 patients that were colonised or developed an infection for part of the hospital stay. While these do not provide an estimate of prevalence (since many of the colonised or infected patients were only colonised for part of the hospital stay) these figures provide upper bounds to the prevalence of colonisation and infection in the study hospital: 9.7% amongst all study patients and 10.2% after exclusions. Putting this into the above formula gives an upper bound for the within-hospital reproduction number of approximately 0.55.

Unlike the study cited above, most studies focus on symptomatic patients and do not test asymptomatic patients at admission. However, the proportion of patients diagnosed with a *C. difficile* infection that were admitted for a *C. difficile* infection (principal diagnosis) is routinely reported. As patients admitted with asymptomatic colonisation who subsequently develop symptoms will not have *C. difficile* infection as their principal diagnosis, this proportion is a lower bound for the total proportion of infections that are due to exposure prior to admission and thus let us estimate an upper bound for the reproduction number. In the USA in the years 1993-2014, 20-34% of admissions who had a *C. difficile* infection had it as their primary diagnosis [31]. This is in excess of typical prevalence of colonisation and infection amongst hospitalised patients: a review of colonisation prevalence reported a range of 4-29% [32]. Therefore, our upper bound for the reproduction number lies in the range 0.69-1.1.

So far, we have assumed hospitalised patients are homogeneous, but this is not the case. Patients who have recently been administered antibiotics are not more susceptible to colonisation but are more likely to develop symptoms and be more infectious [27]. Thus, an SIS model with heterogeneous infectiousness is perhaps more appropriate. However, heterogeneity in infectiousness alone does not affect the estimate of the reproduction number (Section 4.2). Factors affecting susceptibility to colonisation exist and adjusting for these will increase our estimate of the reproduction number (Section 4.1), but this is unfortunately beyond the scope of this case study. However, our simple estimates of $R_0$ are in agreement with more sophisticated models of *C. difficile* transmission in hospitals that have found that the reproduction number is likely to be less than one in many or most hospital settings [33,34].

There are many strains and types of *C. difficile* and it has been suggested that certain strains or types, such as NAP1/RT027, are particularly hospital-adapted [35,36]. It is possible that these strains have significantly higher reproduction numbers in the hospital than we have estimated above and thus may be self-sustaining in hospitals. Unfortunately, we do not have strain-level or type-level data for all strains or types. However, the article used to calculate our first estimate of the reproduction number report the proportion of infections and colonisations typed as NAP1/RT027 [27]. As the authors did not type all isolates, we assume that un-typed isolates were equally likely to be NAP1/RT027 as the isolates from similar patients that were typed, and that the proportion of NAP1/RT027 infections in those with onset <72h after admission (not reported) was similar to patients with colonisation at admission (13%). Under these assumptions, approximately 32 out of 150 (21%) colonisations or infections with NAP1/RT027 were present at admission. Of the approximately 10% of cases that were colonised or infected for some part of their hospital stay, approximately 3% were with NAP1/RT027 and the remaining 7% were with other types. Though colonisation with non-toxigenic strains appears to be protective against infection with toxigenic strains [37], we do not have good information about the interaction of *C. difficile* types. Nevertheless, we can use the argument we presented in section 5 to bound the invasion reproduction number. This becomes

$$R_{\text{Invasion}}^{\text{NAP1}} \leq \frac{1 - \text{Proportion of NAP1 colonisations and infections acquired prior to admission}}{1 - \frac{\text{Prevalence of NAP1 colonisation and infection}}{\text{Proportion } C.difficile \text{ negative} + \text{Prevalence of NAP1 colonisation and infection}}}$$

$$\approx \frac{1 - 0.21}{1 - \frac{0.03}{0.9 + 0.03}}$$

$$\approx 0.8.$$

The basic reproduction number (i.e. in a completely susceptible population without competition with other types) is slightly greater:

$$R_0^{\text{NAP1}} \leq \frac{1 - \text{Proportion of NAP1 colonisations and infections acquired prior to admission}}{1 - \text{Prevalence of any } C.difficile \text{ colonisation and infection}}$$

$$\approx \frac{1 - 0.21}{1 - 0.1}$$

$$\approx 0.9.$$

This suggests that even if other strains were eliminated and NAP1/RT027 did not compete for hosts, the continual importation of colonised and infected individuals would be required to sustain endemic disease in the study hospital. If we perform the same analysis for the pooled non-NAP1/RT027 strains in the study (approximately 212 of 334 colonisations and infections were present on admission) the equivalent upper bounds for the invasion reproduction number and basic reproduction number are both approximately 0.4. Therefore it appears that NAP1/RT027, though importation-driven, was better adapted for transmission in the study hospital than other strains.

## 6.2 *C. difficile* and Animal Reservoirs

Carriage of *C. difficile* in the general adult population is less common than in hospitals or aged-care facilities, with reported prevalence in the range 0-15%, though ≲ 5% is perhaps most typical [32]. *C. difficile* is also commonly found colonising pets and livestock, while *C. difficile* spores are frequently isolated on meat, fresh produce and in water [21]. Crucially, there is significant overlap in strains observed in human and non-human sources [35]. However the proportion of human cases that are acquired from a non-human reservoir is unknown. Consequently, we cannot use our methods to estimate the reproduction number, but we can calculate the reservoir-driven threshold. If it is reasonable to suspect that reservoir exposure accounts for a proportion equal to or exceeding the threshold, then *C. difficile* may be sustained in the human population by exposure to animal reservoirs.

If we begin with a homogeneous SIS model with reservoir exposure, then our estimate of the reservoir-driven threshold is simply the prevalence in the community which is typically ≲ 5% for adults (Section 2). Given the ubiquity of non-human exposure it is plausible that reservoir exposure exceeds this very low threshold. Some individuals will have higher exposure to these reservoirs (depending on diet and lifestyle factors), but this alone will not affect the reservoir-driven threshold unless those with greater exposure are also are more (or less) infectious (Section 4.3). If we heterogeneous infectiousness of those with and without symptoms, or with and without recent antimicrobial exposure, this also does not affect the food driven exposure in isolation (Section 4.2). However, communities are not homogeneous with regards to *C. difficile* colonisation risk, as demonstrated by the higher rates of colonisation and infection in hospitals, aged-care facilities and the very high colonisation rates amongst infants. Accounting for this heterogeneity will increase our estimate of the reservoir-driven threshold (Section 4.1).

If we split our population into four risk categories – (A) hospitalised patients, (B) aged-care residents, (C) infants under 12 months and (D) the rest of the population – we can begin to account for some of this heterogeneity. If we assume separable mixing with heterogeneous susceptibility and infectious period, we need only the prevalence in each group and the proportion of the population that is in each group to estimate the reservoir-driven threshold (equation **7**). The reported range of colonisation prevalence in each of these groups is (A) 0-29%, (B) 0-51%, (C) 18-90% and (D) 0-15% respectively [32], while the total proportion of the population in each of these groups in a developed country like Australia is (A) <0.5% [38], (B) <1% [39], (C) <1.5% [40] and (D) >97% respectively.

If we use the upper end of the prevalence range for each risk group, though only 16.6% of the population is colonised, the reservoir-driven threshold is 48.0%. Assuming a lower colonisation prevalence in the majority population (D) decreases overall prevalence but increases heterogeneity and can increase the reservoir-driven threshold. If only 1% of the healthy adult population is colonised, then overall prevalence is 3.0% but the reservoir-driven threshold is much higher at 81.1%. These extreme values taken from across the literature are not typical and are unlikely to coincide in a single population. If we consider more typical values of colonisation prevalence, the picture is quite different. With prevalence half of the maximum reported values (i.e. (A) 14.5%, (B) 25.5%, (C) 45% and (D) 7.5%), which is still probably much higher than typical for infants in particular [41], the reservoir-driven threshold is only 13.0%. The reservoir-driven threshold is lower still if prevalence is lower in any of the high-risk minority groups (A-C). Figure 3 explores the effect of different prevalence assumptions on the reservoir-driven threshold.

This model and estimate of the reservoir-driven threshold is of course very rough. Transmission is not well mixed between or within the four risk-categories. Furthermore, the pathogen's interactions with medications, gut-flora and host immunity leads to greater complexity than can be captured with a simple SIS model. The risk-categories of individuals change over time as patients age or move in and out of hospitals and so a multi-patch with age structure would provide better estimates. Nevertheless, this very simple calculation serves as a back-of-the-envelope estimate for the plausible range of the reservoir-driven threshold, demonstrating that under a range of reasonable assumptions a relatively small amount of transmission from animals could be sustaining endemic disease in human populations. Our simple calculations with figures from the middle of the reported prevalence range agree with a detailed, model of hospitals and communities that found the reservoir-driven threshold was between 3.5% and 26.0% for a wide range of plausible assumptions.

There are many strains or types of *C. difficile* that circulate in human populations and the arguments set out in section 5 can be used to determine whether individual or types are reservoir-driven. It could be the case that some strains are sustained by exposure to animals, while other strains – though also present in animal populations – are sufficiently transmissible between humans to persist without transmission from animals. *C. difficile* PCR ribotype 078 (RT078) is a particularly good candidate to consider as a reservoir-driven strain. Though it is not known what proportion of human RT078 cases can be attributed to transmission from an animal source, whole-genome sequencing of isolates of this strain from livestock and humans strongly suggest frequent transmission between these groups [42]. On the other hand NAP1/RT027 which is found in livestock but appears to be more transmissible between people than other strains, might have some human cases attributable to animal sources but is less likely to be animal-driven [43]. Finally RT001, which accounts for many human infections in European settings, appears to be uncommon in livestock [43].

# 7 Conclusion

We have outlined the theory and application of very simple rules to estimate reproduction numbers in the presence of reservoir-exposure or imported cases. The rules require minimal information about the population and the pathogen of interest and could be a useful starting point or alternative to more complex models tailored to a population or pathogen. Churcher et al. have developed a statistical test using branching process theory to infer whether $R_0 < 1$ in a population nearing disease elimination but with many imported cases [9]. Cauchemez et. al use a similar approach that accounts for incomplete case detection and the overrepresentation of larger outbreaks to estimate the reproduction number for emerging zoonoses [5]. However, their models assume almost all the population is susceptible and so are not suitable for situations where the prevalence of infection or immunity is far from zero. Moreover, the latter method assumes that the reproduction is less than one so is not appropriate in settings where is there is genuine uncertainty as to whether the reproduction number is above or below one [5]. Our model accounts for susceptible depletion and works for infections where the reproduction number is above or below one, but relies on estimates of prevalence to do so. This can pose a potential difficulty as incidence rather than prevalence is usually reported. Reliable estimates of prevalence either requires near-perfect case acquisition or surveys with large sample sizes especially when prevalence is low. Indeed a good deal of the variability in colonisation prevalence reported for *C. difficile* outside hospitals might be attributed to the relatively small sample sizes involved [32].

Some caution is required when using the reservoir-driven and importation-driven thresholds. It does not follow that if a disease is reservoir-driven or importation driven, then interventions targeting the external source and transmission from the external source will be most effective or 'best'. The 'best' control strategy will depend on the relative effort required to prevent each kind of exposure, the impact of these interventions and metric used to compare these. If it is equally feasible and desirable to eliminate all (or most) exposure from either source, eliminating transmission from the reservoir or importation is clearly the better choice for a reservoir-driven or importation disease as this will prevent all local human cases, while preventing all person-to-person transmission will prevent only the proportion of human cases spread locally by humans. However, if only modest reductions are feasible, then targeting local human transmission may be more effective. One can calculate the normalised derivatives of equilibrium prevalence to estimate the reduction in prevalence achieved by a small reduction in person-to-person transmission or exposure to the external source. For example, in the homogenous SIS model with reservoir-exposure, a greater reduction in prevalence is achieved by reducing person-to-person transmission whenever less than half of cases are acquired from the reservoir [1]. This is true whether or not the disease is reservoir-driven. A similar rule can be derived for the SIS model with imported cases.

The major limitation of our method is the assumption that the disease and population are at equilibrium. Many diseases, including our case study disease *C. difficile,* exhibit seasonal variation [44]. It is possible that an infection is sufficiently transmissible to be locally sustained in high-transmission seasons, but reservoir-driven or importation-driven in low-transmission seasons [9]. Similarly, it possible that exposure to the reservoir is seasonal [45]. It is possible that an epidemic in one setting is driven by exposure to a population or reservoir where an epidemic is ongoing. Our model does not account for these kinds of temporal variability when estimating reproduction numbers and reservoir-driven thresholds.

---

[1] For this simple model the normalised derivatives w.r.t the person-to-person transmission rate and reservoir exposure rate can be written $\frac{\beta}{I}\frac{\partial I}{\partial \beta} = \frac{1-\pi}{\frac{I}{1-I}+\pi}$ and $\frac{f}{I}\frac{\partial I}{\partial f} = \frac{\pi}{\frac{I}{1-I}+\pi}$. Hence $\frac{\beta}{I}\frac{\partial I}{\partial \beta} > \frac{f}{I}\frac{\partial I}{\partial f}$ whenever $\pi < 1/2$.

The simplicity, minimal data requirements, generality and extensibility of the method we have presented here make it useful starting point for understanding the impact and interaction of transmission sources both internal and external to a population.

# 8  Acknowledgements

AM is supported by an Australian Government Research Training Program Scholarship. We thank Laith Yakob for suggested reading that proved valuable for framing our analysis of models with imported cases.

# Figures

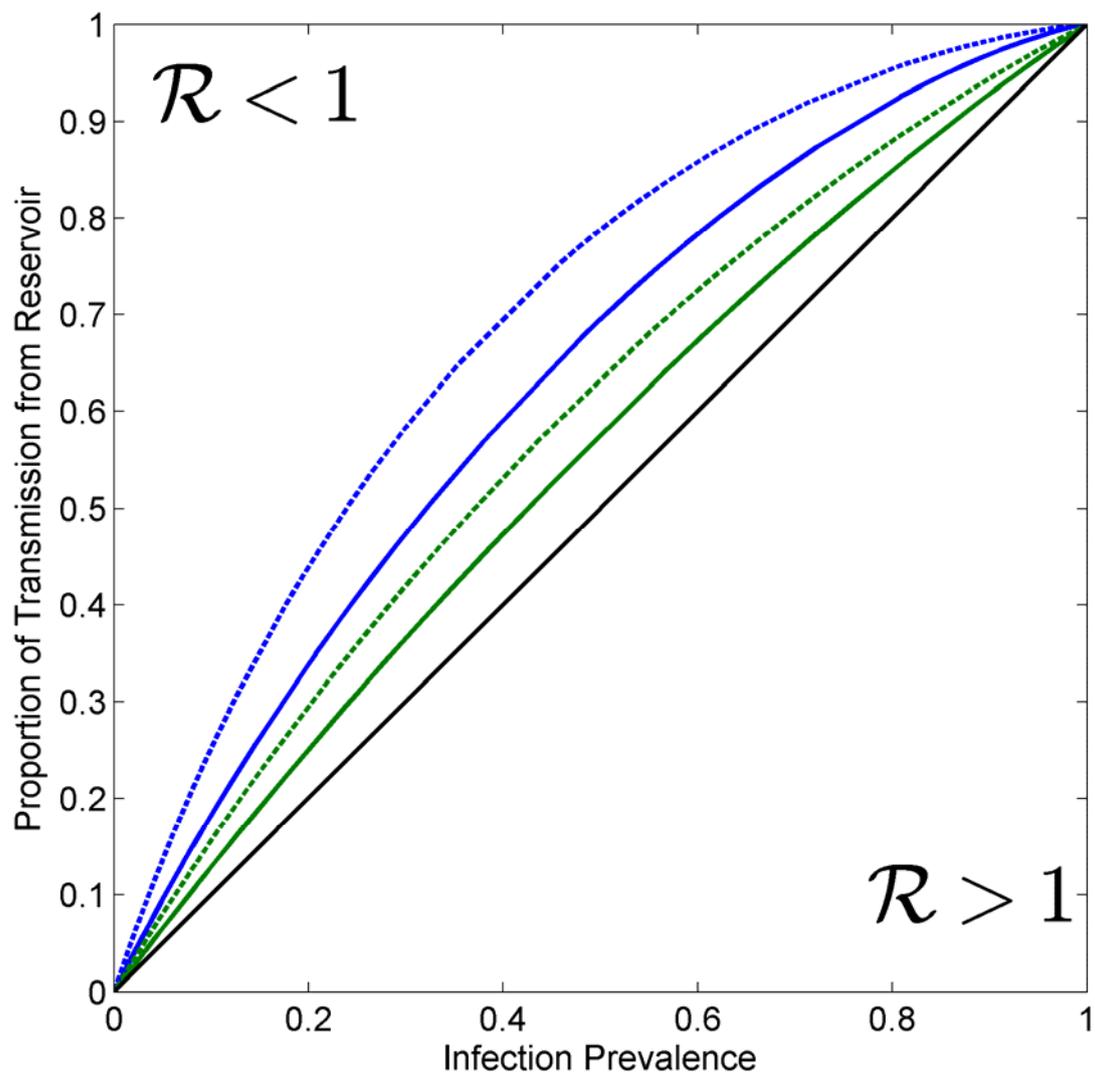

**Figure 1** The reservoir-driven threshold (RDT) – the minimum proportion of transmission attributable to the reservoir above which the basic reproduction number is <1 – as a function of disease prevalence. Each curve indicates the RDT for different population heterogeneity assumptions for infectiousness ($\beta$) and the product of susceptibility and infectious period ($\phi \coloneqq a/\gamma$). The RDT for a homogenous population is equal to the disease prevalence (black line). Heterogeneous $\beta$ alone does not change the RDT (black line). The RDT is higher if $\phi$ heterogeneous and $\beta$ homogenous (solid curves). The size of the effect increases with increasing heterogeneity (green curves: $\phi \sim \Gamma(3, \mu)$, blue curves: $\phi \sim \Gamma(1, \mu)$). Heterogeneity in $\beta$ interacts with heterogeneity in $\phi$, further increasing the RDT if $\beta \propto \phi$ (dashed curves) but decreasing the RDT if $\beta \propto 1/\phi$ (black line).

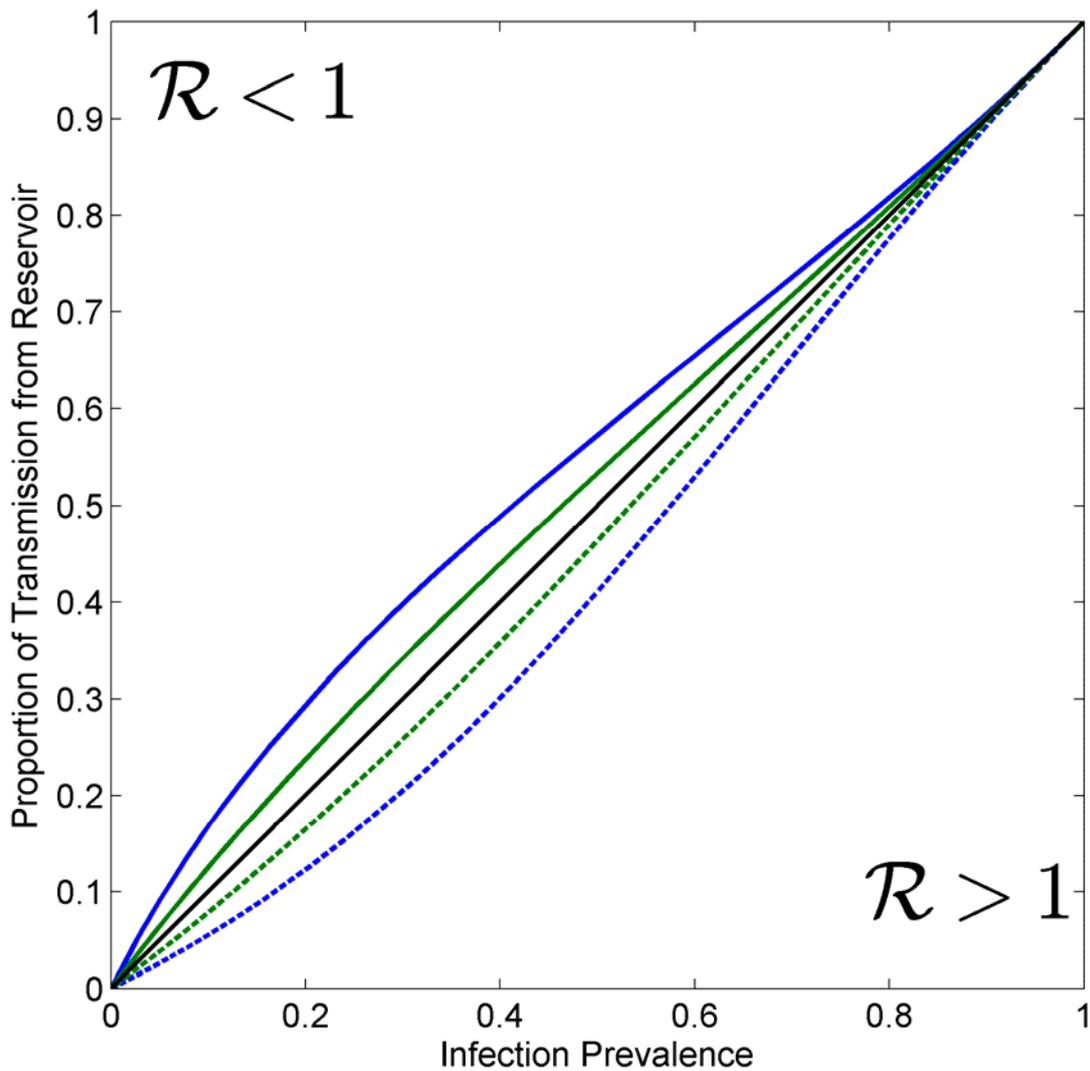

**Figure 2** The reservoir-driven threshold (RDT) for different assumptions for heterogeneity of reservoir exposure ($f$) and person-to-person transmission ($\beta$) across the population. The RDT for a homogenous population is equal to the disease prevalence (black line). The RDT does not change if only $f$ or only $\beta$ is heterogeneous (black line). The RDT is lower if both are heterogeneous and $\beta \propto f$ (dashed curves). The RDT is higher if $\beta$ decreases with increasing $f$ (solid curves: $\beta \propto e^{-f}$). The size of the effect increases with increasing heterogeneity (green curves: $f \sim \Gamma(3, \mu)$, blue curves: $f \sim \Gamma(1, \mu)$).

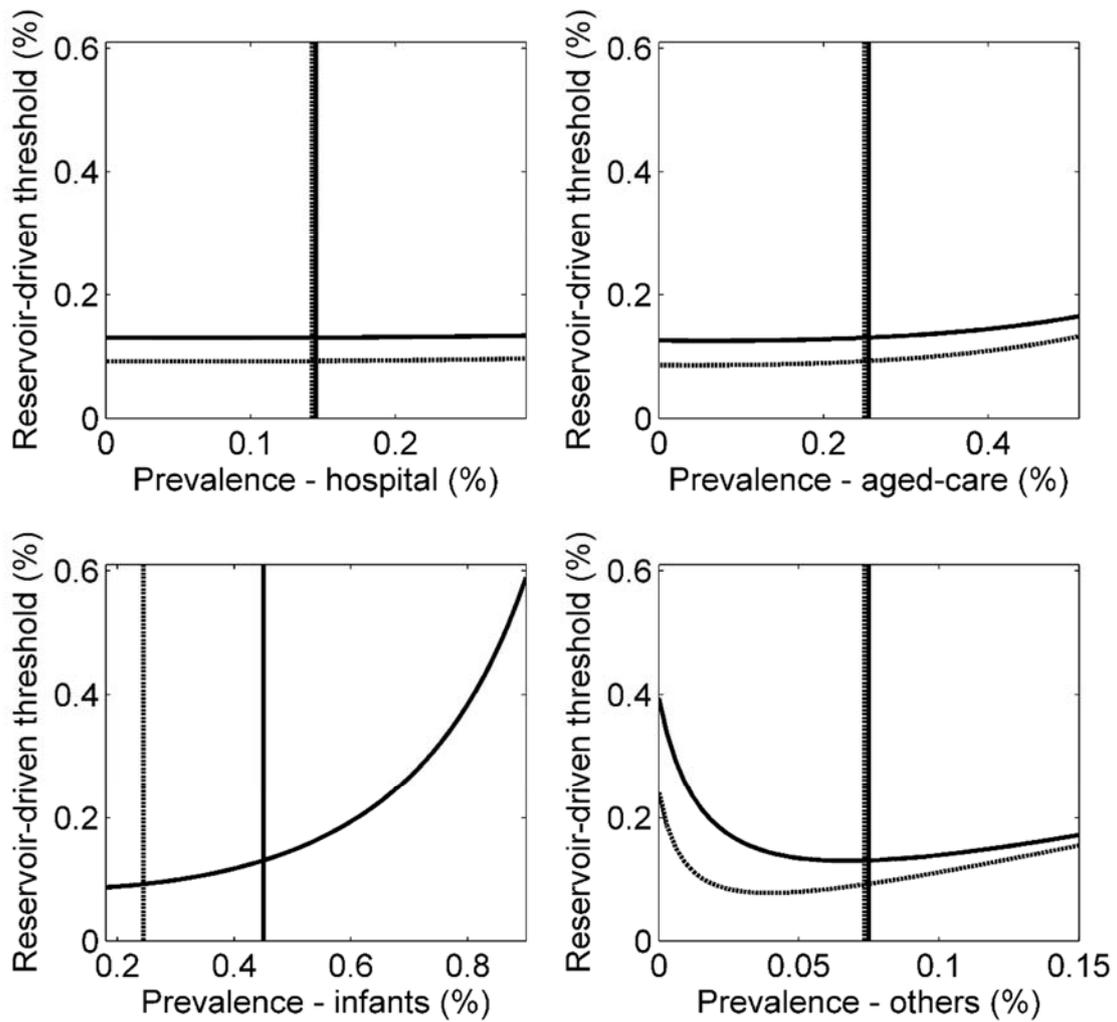

**Figure 3** Estimates of the reservoir-driven threshold for C. difficile in human populations and its dependence on the prevalence of each of four risk groups. In each subfigure, the prevalence in one risk group is varied across the reported range [32] (x-axes) while the other three prevalences are fixed at the values indicated by the vertical lines in the other subfigures. We consider two scenarios; one where each of the fixed prevalences is assumed to be in the middle of the reported range (solid lines and curves); the other the same except the prevalence in infants is only 25% (dotted lines and curves). We assume that 0.5%, 1%, 1.5% and 97% of the population are in the hospital, aged-care, infant and 'other' risk groups respectively.